\documentclass[prd,showkeys,floatfix,nofootinbib,
               preprint,12pt,
               tightenlines,fleqn]{revtex4} 

\usepackage{amsmath,amssymb,revsymb,graphicx,dcolumn}
\usepackage{leftidx}
\usepackage{slashed}
\usepackage[yyyymmdd,hhmmss]{datetime}
\usepackage[ngerman, num]{isodate}
\usepackage{fancyhdr}
\usepackage{verbatim}
\usepackage[usenames, dvipsnames]{color}
 
\definecolor{new}{rgb}{0.858, 0.188, 0.478}

\newcommand{\beq}{\begin{equation}}
\newcommand{\eeq}{\end{equation}}
\newcommand{\beqa}{\begin{eqnarray}}
\newcommand{\eeqa}{\end{eqnarray}}
\newcommand{\bsubeqs}{\begin{subequations}}
\newcommand{\esubeqs}{\end{subequations}}

\usepackage{mwe}

\begin{document}
\hfill KA--TP--02--2017\vspace*{8mm}\newline
\title[]
       {Quantum kinetic theory of a massless scalar model \\ in the presence of a Schwarzschild black hole}
\author{Slava Emelyanov}
\email{viacheslav.emelyanov@kit.edu}
\affiliation{Institute for Theoretical Physics,\\
Karlsruhe Institute of Technology (KIT),\\
76131 Karlsruhe, Germany\\}

\begin{abstract}
\vspace*{2.5mm}\noindent
We employ quantum kinetic theory to investigate local quantum physics in the background of spherically
symmetric and neutral black holes formed through the gravitational collapse. For this purpose in mind,
we derive and study the covariant Wigner distribution function $\mathcal{W}(x,p)$ near to and far away
from the black-hole horizon. We find that the local density of the particle number is negative in the
near-horizon region, while the entropy density is imaginary.
These pose a question whether kinetic theory is applicable in the near-horizon region. We
elaborate on that and propose a possible interpretation of how this result might nevertheless be
self-consistently understood.
\end{abstract}



\maketitle

\section{Introduction}

The black-hole evaporation~\cite{Hawking} is a remarkable discovery in the black-hole physics. This
effect reveals itself in the positive flux of energy density being measurable sufficiently far away from the event
horizon. The thermal profile of the mode distribution characterising this energy flux might imply that one
could define other \emph{local observables} which are usually attributable to normal/classical many-particle
systems (rare gases, plasmas and so on). To our knowledge, there has been no progress in this direction.
In this paper, we shall try to make it by exploiting quantum kinetic theory.

Our main purpose in this paper is therefore to study local kinetic state variables in the background
of evaporating spherically symmetric black holes. A part of the local macroscopic variables correspond 
to the elements of the renormalised stress tensor $\langle \hat{T}_\nu^\mu(x) \rangle$ associated with
a certain field model. These are the energy density, its flux and the pressure. In principle, these do not
need any reference to the kinetic theory as being quantities directly computable from the first principles.
However, there are many other variables which are not. These are the particle density $n(x)$ and the
particle density flux $\mathbf{N}(x)$ as well as the local entropy density $s(x)$ and the entropy density
flux $\mathbf{S}(x)$.

The framework within which we shall be working below is based on a massless scalar field conformally
coupled to gravity. To do quantum kinetic theory, we need the scalar 2-point function $W(x,x')$. It is in
general a difficult problem to analytically derive $W(x,x')$ in Schwarzschild space. Nevertheless, if one
employs the conformal symmetry of the scalar model, one can compute an approximate expression
of the Wightman function close to and far away from the black-hole horizon~\cite{Page}. However, the
2-point function is up to now unknown for physical black holes, i.e. those which have formed through
the gravitational collapse. This is a technical problem we shall analytically address in this paper. The
basic structure of $W(x,x')$ has been already conjectured by us in~\cite{Emelyanov-16b} 
(with further applications in~\cite{Emelyanov-16c}) by exploiting the results of \cite{Page,Candelas}.
We prove this
conjecture in Sec.~\ref{sec:sfm} and derive the higher-order corrections to that result as well. These
corrections turn out to play a crucial role in re-obtaining the correct expression of the renormalised
stress-energy tensor.

These preliminary steps will allow us to derive a covariant Wigner distribution $\mathcal{W}(x,p)$, 
wherein $x$ and $p$ denote a space-time point and four-momentum (e.g.,
see~\cite{deGroot&vanLeeuwen&vanWeert}). We then apply this distribution for the derivation of the
local macroscopic variables in the region far away and near to the black-hole horizon. Our results are
presented in Sec.~\ref{sec:qkf}. To sum it up, we find the standard picture far away from the event horizon,
whereas its inapplicability in the near-horizon region as $n(x)$ and $s(x)$ turn out to be negative and
imaginary, respectively.

We discuss our results in Sec.~\ref{sec:cr} and propose physical interpretation of how these
can possibly be understood in a self-consistent manner.

Throughout this paper the fundamental constants are set to $c=G=k_\text{B} = \hbar = 1$, unless stated
otherwise. We shall be employing a convention for the indices $``0"$ and $``1"$ to refer to the far-horizon
region and the near-horizon region, respectively. The logic behind this notation is that $r_H/r \rightarrow 0$
and $r_H/r\sim 1$ hold far away from and near to the event horizon.

\section{Scalar field model}
\label{sec:sfm}

We shall be dealing with the scalar field $\Phi(x)$ in the background of Schwarzschild black hole of
astrophysical mass $M$. We set the scalar-field mass to zero and assume the field is conformally
coupled to gravity. The scalar Lagrangian is thus taken to be of the form
\beqa
\mathcal{L} &=& - \frac{1}{2}\Phi\Box\Phi + \frac{1}{12}\,R\Phi^2\,,
\eeqa
where $R$ is the Ricci scalar which is, however, identically zero in the Schwarzschild geometry
described by
\beqa
ds^2 &=& g_{\mu\nu}dx^\mu dx^\nu \;=\; f(r)dt^2 - \frac{dr^2}{f(r)} - r^2d\Omega^2\,,
\eeqa
where the lapse function $f(r) \equiv 1 - r_H/r$ and $d\Omega$ is an element of the solid angle. 
The parameter $r_H \equiv 2M$ is the Schwarzschild radius aka the size of the event horizon.

\subsection{Scalar Wightman function in Schwarzschild frame}

We have recently conjectured the structure of the Wightman 2-point function in the background of
evaporating Schwarzschild black hole~\cite{Emelyanov-16b},
wherein we have also used that to study the one-loop effects in QED in the far-from-horizon region
and in a massless scalar model with the quartic self-interaction term in~\cite{Emelyanov-16c}.
In this section, we shall prove that and also derive higher-order corrections 
with respect to $\Delta\mathbf{x}$ which have been neglected in~\cite{Emelyanov-16b}.

It turns out that the approximate analytic expression of the scalar Wightman function $W(x,x')$ can be found
with a comparably little computational effort if one takes advantage of the conformal symmetry of the
scalar model. This observation allowed to compute the Wightman function for the Hartle-Hawking (HH) as
well as Boulware state~\cite{Page}.

To derive the 2-point function for the physical black holes, we first perform a conformal transformation
of the Schwarzschild metric to its ultra-static form
$\bar{g}_{\mu\nu}(x)$, namely
\beqa
g_{\mu\nu}(x) &=& f(r)\,\bar{g}_{\mu\nu}(x)\,.
\eeqa
Correspondingly, the Wightman function $W(x,x')$ fulfilling $\Box W(x,x') = 0$ (we have taken into
account here that $R = 0$ for the Schwarzschild black hole) can be written in the ultra-static
metric as follows
\beqa
W(x,x') &=& \bar{W}(x,x')/\big(f(r)f(r')\big)^{\frac{1}{2}}\,,
\eeqa
where $\bar{W}(x,x')$ satisfies
\beqa\label{eq:gi-sp:equation}
\Big(\bar{\Box} - \frac{1}{6}\bar{R}\Big)\bar{W}(x,x') &=& 0\,.
\eeqa
According to our convention being extensively used below, all barred quantities are
defined with respect to the ultra-static metric $\bar{g}_{\mu\nu}(x)$.

Since the Killing algebra of the space under consideration consists of the time translation as well
as three generators of the rotational group, we look for a solution of~\eqref{eq:gi-sp:equation} 
in the form
\beqa\label{eq:gi-sp:ansatz}
\bar{W}(x,x') &=& \int_\mathbf{R}\frac{d\omega}{4\pi\omega}\,
e^{-i\omega(t-t')}\bar{K}_\omega(\mathbf{x},\mathbf{x}')\,,
\eeqa
where the integral is over $\omega \in (-\infty,+\infty)$ and by definition
\beqa\label{eq:gi-sp:ansatz-2}
\bar{K}_\omega(x,x') &\equiv& \frac{1}{4\pi}\bar{\Delta}^{\frac{1}{2}}(\mathbf{x},\mathbf{x}')\,
\frac{\sin(\omega\rho)}{\omega\rho}\,\chi_\omega(\mathbf{x},\mathbf{x}') \quad
\text{with} \quad \mathbf{\rho} \;\equiv\; \big(2\bar{\sigma}(\mathbf{x},\mathbf{x}')\big)^{\frac{1}{2}}\,,
\eeqa
where $\bar{\sigma}(\mathbf{x},\mathbf{x}')$ is a geodetic interval for the spatial section of the
ultra-static metric $\bar{g}_{\mu\nu}(x)$ and $\bar{\Delta}(\mathbf{x},\mathbf{x}')$ the Van Vleck-Morette
determinant.

Substituting~\eqref{eq:gi-sp:ansatz} into~\eqref{eq:gi-sp:equation}, one obtains an equation which
the unknown bi-scalar $\chi_\omega(\mathbf{x},\mathbf{x}')$ must satisfy. Specifically, this reads
\beqa\label{eq:eq-for-bi-scalar-chi}
\bar{\Box}\chi_\omega 
+ \frac{1}{3}\big(\bar{R}_i^j - 2\omega^2\delta_i^j \big)\bar{\sigma}^i\bar{\nabla}_j\chi_\omega
-\frac{1}{12}\big(2\bar{R}_{ik;j} - \bar{R}_{ij;k}\big)\bar{\sigma}^i\bar{\sigma}^j\bar{\nabla}^k\chi_\omega
+ \text{O}\big((\bar{\sigma}^i)^3\big) &=& 0\,,
\eeqa
where $\bar{\sigma}^i \equiv \bar{\nabla}^i\bar{\sigma}$ and $i,j$ run from $1$ to $3$. In the 
derivation of Eq.~\eqref{eq:eq-for-bi-scalar-chi} we have employed the fact that
$\bar{R}_{;i} - 2\bar{R}_{i;j}^j$ identically vanishes for \emph{any} lapse function $f(r)$ and
\beqa
9\bar{R}_{;ij} + 9\bar{R}_{ij;k}^{\;\;\;;k} - 24\bar{R}_{ik;j}^{\;\;\;;k}
-12\bar{R}_{ik}\bar{R}_j^k + 6\bar{R}^{kn}\bar{R}_{ikjn} +
4\bar{R}_{iknm}\big(\bar{R}_{j}^{\;\;mnk} + \bar{R}_j^{\;\;knm}\big) &=& 0
\eeqa
for the lapse function of the form $1 - r_H/r + \Lambda r^2/3$. It means that the equation
\eqref{eq:eq-for-bi-scalar-chi} is applicable to a wide class of static spacetimes. As a consequence,
we have
\beqa
\bar{\Box}\bar{\Delta}^\frac{1}{2}(\mathbf{x},\mathbf{x}') &=& 
\frac{1}{6}\,\bar{R}(\mathbf{x})\bar{\Delta}^\frac{1}{2}(\mathbf{x},\mathbf{x}')
+ \text{O}\big((\bar{\sigma}^i)^3\big)\,.
\eeqa

We need now to solve Eq.~\eqref{eq:eq-for-bi-scalar-chi} to eventually obtain the 2-point function.
In the leading order of the approximation, the bi-scalar $\chi_\omega(x,x')$ reads
\beqa\label{eq:bi-scalar}
\chi_\omega(\mathbf{x},\mathbf{x}') &=& a_\omega + b_\omega\frac{(f(r)f(r'))^{\frac{1}{2}}}{rr'}
\Big(1 \pm i\omega\Delta{r}_\star + \alpha_{\omega}(r,r')\Delta{r}_\star^2
+ \beta_{\omega}(r,r')\bar{\sigma}(\mathbf{x},\mathbf{x}')\Big),
\eeqa
where $\Delta{r}_\star \equiv r_\star - r_\star'$ with $r_\star$ denoting the Regge-Wheeler radial
coordinate and
\bsubeqs\label{eq:fun-coefficients}
\beqa
\alpha_\omega(r,r') &\approx& - \frac{\omega^2}{2}\,,
\\[1mm]
\beta_\omega(r,r') &\approx& + \frac{\omega^2}{3}+\frac{r_H}{12(rr')^\frac{3}{2}}
+ \frac{r_H}{4(rr')^\frac{3}{2}}\big(f(r)f(r')\big)^\frac{1}{2}\,.
\eeqa
\esubeqs
The sign in front of the second term in the parenthesis of Eq.~\eqref{eq:bi-scalar} cannot be fixed without
referring to the mode expansion of the scalar field. We take it negative in the far-horizon region and positive
in the near-horizon region, because then $\langle \hat{T}_t^r\rangle$ has a correct sign.

The bi-scalar $\chi_\omega(\mathbf{x},\mathbf{x}')$ given in \eqref{eq:bi-scalar} with
\eqref{eq:fun-coefficients} is a solution of the equation $\bar{\Box}\chi_\omega(\mathbf{x},\mathbf{x}') = 0$
up to the order of $\text{O}(\mathbf{x}-\mathbf{x}')$. There are infinitely many solutions of this type.
However, we shall show below that the bi-functions defined in Eq.~\eqref{eq:fun-coefficients} yield the
stress-energy tensor of the scalar field as found in~\cite{Christensen&Fulling,Candelas}.\footnote{There are
extra terms in $\alpha_\omega(r,r')$ and $\beta_\omega(r,r')$ which give sub-leading
contributions to the diagonal elements of $\langle \hat{T}_\nu^\nu\rangle$ in both far-horizon and
near-horizon region. For instance, there are additional terms vanishing as $f(r)$ near the horizon which we have omitted. We shall study these in detail elsewhere.}

We now need to determine the functions $a_\omega$ and $b_\omega$. With this purpose in mind, 
we consider the far-horizon ($r,r' \gg r_H$) and near-horizon ($r,r' \sim r_H$) region separately.

\subsubsection{Far-horizon region}

One might expect from the physical grounds that the Wightman function $W(x,x')$ must reduce to
the Minkowski 2-point function, $W_M(x,x')$, in the asymptotically flat region, i.e. 
$W(x,x') \rightarrow W_M(x,x')$ in the limit $r \rightarrow \infty$ with $|r-r'| \ll R$, where $R$ is the
distance to the black-hole centre here and below. Indeed, if it had
not be the case, then it would be not legitimate to use the Minkowski-space approximation in
describing and testing particle physics in the colliders on earth. This implies 
\beqa
a_{\omega,0} &\approx& +4\omega^2\,\theta(+\omega)\,,
\eeqa
where $\theta(z)$ is the Heaviside step function. 

The function $b_{\omega,0}$ cannot be so simply determined. However, if we set
\beqa
b_{\omega,0} &\approx& +27\,(\omega M)^2n_\beta(\omega)\big(\theta(\omega)
+ e^{\beta\omega}\theta(-\omega)\big)
\quad \text{with} \quad n_\beta(\omega) \;\equiv\; 1/(e^{\beta\omega} - 1)\,,
\eeqa
where $\beta = 2\pi/\kappa \equiv 8\pi M$ is the inverse Hawking temperature
$T_H \equiv 1/8\pi M$~\cite{Hawking}, then we obtain our previous
result~\cite{Emelyanov-16b,Emelyanov-16c} (with $\chi_\omega(\mathbf{x},\mathbf{x}')$ in the
limit $\mathbf{x}' \rightarrow \mathbf{x}$) which is in agreement with~\cite{Candelas,Page}.
Specifically, we find
\beqa\label{eq:2pf-far}
W_0(x,x') &=& W_M(x,x') + \Delta{W}_0(x,x')\,,
\eeqa
where the first term is the Minkowski 2-point function, i.e.
\beqa
W_M(x,x') &\approx& {\int}\frac{d^3\mathbf{k}}{(2\pi)^3}\frac{1}{2k_0}\,\exp(-ik\Delta{x})
\quad \text{with} \quad k_0 \;=\;|\mathbf{k}|\,,
\eeqa
and the higher-order correction to $W_M(x,x')$ reads
\beqa\label{eq:2pf-far-correction}
\Delta W_0(x,x') &\approx& \frac{27r_H^2}{16R^2}
{\int}\frac{d^3\mathbf{k}}{(2\pi)^3}\frac{n_\beta(k_0)}{k_0}\,\cos(\mathbf{k}\Delta\mathbf{x})
\\[1mm]\nonumber
&& \quad\quad\quad \times
\left[\Big(1 - \frac{k_0^2}{6}\big(3\Delta{r}^2 - \Delta\mathbf{x}^2\big)\Big)\cos(k_0\Delta{t})
+k_0\Delta{r}\sin(k_0\Delta{t})\right],
\eeqa
where $k\Delta{x} \equiv k_\mu\Delta{x}^\mu$ and
$\Delta\mathbf{x}^2 \equiv \Delta{x}^2 + \Delta{y}^2 + \Delta{z}^2$ by our convention. The
coordinates $x$, $y$ and $z$ here are local Cartesian coordinates introduced at the
distance $R$ from the centre of the black hole.

The 2-point function~\eqref{eq:2pf-far} is more general than that we have found
in~\cite{Emelyanov-16b,Emelyanov-16c} as it also contains the higher-order corrections
in $\Delta\mathbf{x}$. We shall show below that $\Delta W_0(x,x')$ yields the correct expression
of the renormalised stress tensor $\langle \hat{T}_\nu^\mu \rangle$ at the spatial infinity.

\subsubsection{Near-horizon region}

One might expect from the equivalence principle that the Wightman function $W(x,x')$ in the near-horizon
region $r,r' \sim r_H$ with $|r-r'| \ll r_H$ must approximately be given by the 2-point function $W_M(x,x')$
as in Minkowski space when expressed in the \emph{local inertial coordinates}. We take this for granted
below to fix the function $a_{\omega,1}$. 

To determine $a_{\omega,1}$, one needs first to introduce new spatial coordinates instead of the
angle coordinates in the following manner: 
\beqa
y^2 + z^2 &=& 4r_H^2\tan^2(\theta/2) \quad \text{and} \quad z/y \;=\; \tan\phi
\quad \text{with} \quad y^2 + z^2 \;\ll\; r_H^2\,.
\eeqa
The geodetic distance $\sigma(x,x')$ in these coordinates acquires a comparably simple structure, namely
\beqa
\sigma(x,x') &\approx& \frac{1}{\kappa^2}\big(f(r)f(r')\big)^\frac{1}{2}\bigg(\cosh\kappa\Delta{t} - 
\frac{f(r) + f(r') + \kappa^2(\Delta{y}^2 + \Delta{z}^2)}{2\big(f(r)f(r')\big)^\frac{1}{2}}\bigg)\,,
\eeqa
where we have neglected the higher-order corrections and $\kappa \equiv 1/2r_H$ by definition.
It is worth emphasising that $\sigma(x,x')$ has been computed directly as a \emph{geometrical}
quantity in Schwarzschild space. Having calculated the geodetic distance 
$\bar{\sigma}(\mathbf{x},\mathbf{x}')$ of the spatial section of the ultra-static metric
$\bar{g}_{\mu\nu}(x) = g_{\mu\nu}(x)/f(r)$ in the near-horizon region, we then obtain
\beqa
\sigma(x,x') &\approx& \frac{1}{\kappa^2}\big(f(r)f(r')\big)^\frac{1}{2}
\big(\cosh\kappa\Delta{t} - \cosh\kappa\rho\big)\,,
\eeqa
where $\rho \equiv (2\bar{\sigma}(\mathbf{x},\mathbf{x}'))^\frac{1}{2}$ as defined in
Eq.~\eqref{eq:gi-sp:ansatz-2}. Therefore, in order to have $W(x,x')$ be approximately equal to
$-1/(8\pi^2\sigma(x,x'))$ in the near-horizon region, one must set
\beqa
a_{\omega,1} &\approx& +4\omega^2 n_\beta(\omega)
e^{\beta\omega}\big(\theta(+\omega) + \theta(-\omega)\big)\,.
\eeqa

In order to determine the function $b_{\omega,1}$, one needs to treat the mode expansion of the
scalar field operator $\hat{\Phi}(x)$. Employing the results of~\cite{Candelas,Page}, we obtain
\beqa
b_{\omega,1} &\approx& -27(\omega M)^2n_\beta(\omega)\big(
\theta(+\omega) + e^{\beta\omega}\theta(-\omega)\big)\,.
\eeqa

Thus, the 2-point function near the horizon reads
\beqa\label{eq:2pf-near}
W_1(x,x') &=& W_M(x,x') + \Delta{W}_1(x,x')\,,
\eeqa
where $W_M(x,x') \approx W_\text{HH}(x,x')$ for $r,r' \rightarrow r_H$ and
\beqa
W_\text{HH}(x,x') &\approx& \frac{\bar{\Delta}^\frac{1}{2}(\mathbf{x},\mathbf{x}')}{(f(r)f(r'))^\frac{1}{2}}
{\int}\frac{\omega d\omega}{(2\pi)^2}\,n_\beta(\omega)e^{\beta\omega}
\Big[e^{-i\omega\Delta{t}} + e^{-\beta\omega}e^{+i\omega\Delta{t}}\Big]\frac{\sin \omega\rho}{\omega\rho}
\eeqa
with $\omega > 0$, and the correction to $W_M(x,x')$ reads
\beqa\label{eq:2pf-near-correction}
\Delta{W}_1(x,x') &\approx& - \frac{27}{16}\bar{\Delta}^\frac{1}{2}(\mathbf{x},\mathbf{x}')
{\int}\frac{\omega d\omega}{(2\pi)^2}\,n_\beta(\omega)\,\frac{\sin \omega\rho}{\omega\rho}
\\[1mm]\nonumber
&&\;\;\times
\left[\Big(1-\frac{\omega^2}{2}\Delta{r}_\star^2 + \frac{1}{3}(\omega^2 + \kappa^2)\bar{\sigma}(\mathbf{x},\mathbf{x}')\Big)\cos(\omega\Delta{t}) - \omega\Delta{r}_\star\sin(\omega\Delta{t}) \right].
\eeqa
The 2-point function $W_1(x,x')$ is also more general than that we have found
in~\cite{Emelyanov-16b,Emelyanov-16c} in the near-horizon region, but reduces to it if one takes
the limit $\mathbf{x}' \rightarrow \mathbf{x}$ in the square brackets.

\subsection{Scalar Wightman function in Fermi frame}

For the applications below, we need to express the scalar 2-point function $W_1(x,x')$ via the Fermi
normal coordinates. These coordinates are characterised by a geodesic with the tangent vector $G$~\cite{Manasse&Misner}.

One can introduce an orthonormal tetrad $e_a^\mu = (e_{t_F}^\mu, e_{x_F}^\mu, e_{y_F}^\mu, e_{z_F}^\mu)$,
such that $e_{t_F}^\mu$ is a (time-like) tangent vector to the geodesic $G$ describing a \emph{radial free fall}
towards the black hole. This tetrad is given by
\bsubeqs\label{eq:fermi-tetrad}
\beqa
e_{t_F}^\mu\partial_\mu &=& \frac{1}{f(r)}\,\partial_t - \big(1 - f(r)\big)^{\frac{1}{2}}\partial_r\,,
\\[0mm]
e_{x_F}^\mu\partial_\mu &=& - \frac{\big(1 - f(r)\big)^{\frac{1}{2}}}{f(r)}\,\partial_t + \partial_r\,,
\\[1mm]
e_{y_F}^{\mu}\partial_\mu &=& \frac{1}{r}\,\partial_\theta\,,
\quad
e_{z_F}^{\mu}\partial_\mu \;=\; \frac{1}{r\sin\theta}\,\partial_\phi\,.
\eeqa
\esubeqs
In the Fermi normal coordinates $x^a = (t_F,x_F,y_F,z_F)$, the metric tensor along the geodesic has
the Minkowski form, i.e. $g_{ab}|_G = \eta_{ab}$, such that $\Gamma_{ab}^c|_G = 0$. It is worth noticing
that the Fermi time $t_F$ is identical to the Painlev\'{e}-Gullstrand time $\tau$ we have made use
of in~\cite{Emelyanov-16c}. For space-time points close to the geodesic $G$, one has
\beqa\label{eq:fermi-metric}
g_{ab}(x_F) &=& \eta_{ab} - \kappa_{ab} R_{acbd}(x_F)x_F^cx_F^d + \text{O}\big(x_F^3\big)\,,
\eeqa
as shown in~\cite{Manasse&Misner}, where there is no summation over $a$ and $b$
in the second term of~\eqref{eq:fermi-metric} and
\beqa
\kappa_{ab} &=& \frac{1}{2}\Big(\delta_a^0 + \delta_b^0
+ \frac{1}{3}\sum\limits_{i = 1}^3\big(\delta_a^i + \delta_b^i\big)\Big)\,.
\eeqa

The geodetic distance $\sigma(x,x')$ in these coordinates is given by
\beqa
\sigma(x,x') &=& \sigma(x_F,x_F') \;=\;
\frac{1}{2} \eta_{ab}(x_F - x_F')^a(x_F -x_F')^b + \text{O}\Big(\frac{r_H\Delta{x}_F^4}{R^3}\Big).
\eeqa
where we have taken one of the points on the geodesic.

In the local inertial frame associated with the geodesic $G$, the Wightman function $W(x,x')$
should naturally be given by $W_M(x,x')$ whenever it is legitimate to neglect geometrical corrections.
This is exactly what we have obtained in Eq.~\eqref{eq:2pf-far} and Eq.~\eqref{eq:2pf-near}. This means
no significant (quantum) effect can be discovered in the near-horizon region for an observer freely
falling in the black hole of a sufficiently large mass $M$.\footnote{For black holes of mass in the range
$10^{10}\,\text{g} \lesssim M \ll 10^{16}\,\text{g}$,
there are certain tiny effects (the modification of the light deflection angle and Debye-like screening of a
point-like charge) which might be testable~\cite{Emelyanov-16a,Emelyanov-16b} assuming
these exist in nature and their number density is sufficiently large.}

Far away from the black-hole horizon, the Schwarzschild coordinates $x$ go over to the Minkowski
coordinates $x_M$. Therefore, the 2-point function $W_0(x,x')$ is already given in the flat coordinates
for $R \gg r_H$. In the region near to the black-hole horizon, the Schwarzschild coordinates $x$
considerably differ from the Fermi ones $x_F$. Having computed $x_F$ as functions of $x$ from
\eqref{eq:fermi-tetrad}, we obtain
\beqa
W_1(x_F,x_F') &=& W_M(x_F,x_F') + \Delta{W}_1(x_F,x_F')\,,
\eeqa
where
\beqa
\Delta{W}_1(x_F,x_F') &\approx& -\frac{27}{4}{\int}\frac{d^3\mathbf{k}}{(2\pi)^3}\frac{n_{2\beta}(k_0)}{k_0}\,
\cos(\mathbf{k}\Delta\mathbf{x}_F)
\\[1mm]\nonumber
&& \quad\quad \times
\left[\Big(1 - \frac{k_0^2}{6}\big(3\Delta{x}_F^2 - \Delta\mathbf{x}_F^2\big)\Big)\cos(k_0\Delta{t}_F)
- k_0\Delta{x}_F\sin(k_0\Delta{t}_F)\right],
\eeqa
where $\mathbf{x}_F = (x_F,y_F,z_F)$ by definition. It should be mentioned that we have explicitly checked
that $\Delta{W}_1(x_F,x_F')$ coincides with
$\Delta{W}_1(x,x')$ given in Eq.~\eqref{eq:2pf-near-correction} up to the order of $\Delta{x}_F^2$ including.
Therefore, the above expression of $\Delta{W}_1(x_F,x_F')$ should be reliable up to this order only.
The same holds for the 2-point function $W_0(x,x')$ far away from the black hole. The reason of this
limitation comes from the bi-scalar $\chi_\omega(\mathbf{x},\mathbf{x'})$ which we have determined only
up to that order. This approximation is, however, adequate for our purposes below.

\section{Quantum kinetic approach to black-hole physics}
\label{sec:qkf}

\subsection{Relativistic kinetic theory: Brief introduction}

Many-particle systems can be described with the aid of the local macroscopic state variables. These
variables are the particle density number, energy density, pressure and so on. In the framework of the
kinetic theory, these are defined through the one-particle distribution function. This distribution is
usually denoted by $f_\text{cl}(x,p)$, where $x^\mu = (t,\mathbf{x})$ and $p^\mu = (p_0,\mathbf{p})$ with
$g_{\mu\nu}p^\mu p^\nu = m^2$ are a space-time coordinate and momentum coordinate, respectively.

The space-time evolution of the distribution function $f_\text{cl}(x,p)$ is governed by the transport
equation, which is a relativistic generalisation of the famous Boltzmann equation. Specifically, this
reads
\beqa\label{eq:transport-equation}
\Big(p^\mu\frac{\partial}{\partial x^\mu} - 
\Gamma_{\mu\nu}^\lambda p^\mu p^\nu \frac{\partial}{\partial p^\lambda}
\Big) f_\text{cl}(x,p) &=& C[f_\text{cl}(x,p)]
\eeqa
in curved spacetime, where $C[f_\text{cl}(x,p)]$ is a collision integral taking into account
binary scattering processes of the constituent particles of the system, and no external field apart
from gravity has been assumed (e.g., see~\cite{Cercignani&Kremer}).

The main state variables are the particle four-current, i.e.
\beqa
N^{\mu}(x) &=& (-g)^\frac{1}{2}{\int}\frac{d^3\mathbf{p}}{p_0}\,p^\mu f_\text{cl}(x,p)\,,
\eeqa
the energy-momentum tensor reading
\beqa
T^{\mu\nu}(x) &=& (-g)^\frac{1}{2}{\int}\frac{d^3\mathbf{p}}{p_0}\,p^\mu p^\nu f_\text{cl}(x,p)\,,
\eeqa
and the entropy four-flow defined as
\beqa
S^{\mu}(x) &=& -(-g)^\frac{1}{2}{\int}\frac{d^3\mathbf{p}}{p_0}\,p^\mu f_\text{cl}(x,p)
\big(\ln(h^3f_\text{cl}(x,p))-1\big)\,.
\eeqa
These macroscopic variables are local. This fact allows to describe equilibrium as well
as non-equilibrium macroscopic states of the system by these variables. 

The macroscopic conservation laws
\bsubeqs
\beqa
\nabla_\mu N^\mu(x) &=& 0\,,
\\[1mm] 
\nabla_\mu T^{\mu\nu}(x) &=& 0\,.
\eeqa
\esubeqs
can be shown to hold as a consequence of the transport equation \eqref{eq:transport-equation} and the
microscopic conservation laws of the particle number as well as the four-momentum. These can in
turn be used to derive the well-known hydrodynamic equations like the continuity equation or the Euler
equations with the relativistic corrections. 

In the kinetic theory, one can also prove the Boltzmann $H$-theorem. This theorem states that the entropy
production rate $\sigma(x)$ at any space-time point is never negative, i.e.
\beqa
\sigma(x) &\equiv& \nabla_\mu S^\mu(x) \;\geq\; 0\,.
\eeqa

For the sake of completeness, we want finally to remind about the role of the hydrodynamic (e.g., 
Eckart or Landau-Lifshitz) velocity $U^\mu(x)$. This allows to define covariant state variables corresponding
to the particle number density $N^\mu U_\mu$, the energy density $T_{\mu\nu} U^\mu U^\nu$ and so on.
For more details, we refer to the references~\cite{deGroot&vanLeeuwen&vanWeert,Cercignani&Kremer}.

\subsection{Covariant Wigner distribution $\mathcal{W}(x,p)$}

Starting with local quantum field theory, one can introduce the distribution function and the transport equation
associated with it. This leads to quantum kinetic theory~\cite{deGroot&vanLeeuwen&vanWeert}. The
main object in this theory is the Wigner operator
\beqa\label{eq:wigner-distribution}
\hat{\mathcal{W}}(x,p) &\equiv& \frac{4\pi}{(2\pi)^5}{\int}d^4y\, e^{-ipy}\,\hat{\Phi}\big(x+y/2\big)
\hat{\Phi}\big(x-y/2\big)\,.
\eeqa
Note that we are working in the Fermi coordinate frame here. It allows us to employ the ordinary Fourier
transform as if the Fermi frame is infinitely large. This does not serve any problem whenever the physics
we are interested in is characterised by \textcolor{new}{a} length scale being much smaller than a characteristic curvature
scale.

Once we have a quantum system described by a certain state, we can relate with it the Wigner
distribution
\beqa
\mathcal{W}(x,p) &=& \langle\hat{\mathcal{W}}(x,p)\rangle\,.
\eeqa
It is worth pointing out that there is no direct probabilistic interpretation of the covariant
Wigner distribution in terms of particles~\cite{deGroot&vanLeeuwen&vanWeert}. This can be understood
from its very definition which is entirely independent on the covariant wave function. This is in contrast
to the classical one-particle distribution $f_\text{cl}(x,p)$ introduced above, which is a probability density
giving a number of particles in a spatial volume $\Delta\mathbf{x}^3$ with the momentum in the interval
between $\mathbf{p}$ and $\mathbf{p} + \Delta{\mathbf{p}}$. Despite of the lack of the straightforward 
physical meaning of $\mathcal{W}(x,p)$ in terms of particles, we want to employ $\mathcal{W}(x,p)$
in Schwarzschild geometry in order to get further insights about black holes and physics related to their
evaporation.

\subsection{Wigner distribution in presence of evaporating black hole}

We now derive the covariant Wigner distribution associated with the scalar model in the background
of the black hole formed via the gravitational collapse. 

\subsubsection{Far-horizon region}

Far away from the black-hole horizon $R \gg r_H$, the 2-point function is given by \eqref{eq:2pf-far}.
The contribution to the Wigner distribution \eqref{eq:wigner-distribution} comes from only those
terms in $W_0(x,x')$ which contain the positive frequency modes. The result reads
\beqa\label{eq:wd-far}
\mathcal{W}_0(x,p) &\approx& \frac{3^3M^2n_\beta(p_0)}{2^5\pi^3 R^2p_0}
\left[1 - p_0P_1\Big(\frac{\mathbf{n}\mathbf{p}}{p}\Big)\partial_p
+ \frac{p_0^2}{3p}P_2\Big(\frac{\mathbf{n}\mathbf{p}}{p}\Big)
\big(p \partial_p^2 - \partial_p\big)\right]{\delta(p_0 - p)}\,,
\eeqa
where $P_n(z)$ is the Legendre polynomial of the order $n \in \mathbb{N}_0$ and
\beqa
p &\equiv& |\mathbf{p}| \quad \text{and} \quad \mathbf{n} \;\equiv\; \mathbf{R}/R\,.
\eeqa 
In deriving the Wigner distribution $\mathcal{W}_0(x,p)$, we have omitted terms which
vanish faster than $1/R^2$ in the limit $R \rightarrow \infty$. 

The prefactors of the first, second and third term in $\mathcal{W}_0(x,p)$ are proportional to
the Legendre polynomial of the zeroth, first and second order, respectively. This resembles 
a similar structure of the monopole, dipole and quadrupole potentials in the multiple expansion of the
electrostatic potential performed sufficiently far away from a gas of charged particles. Therefore,
we shall occasionally refer to these terms in the following as to the monopole-, dipole- and
quadrupole-\emph{like} term, respectively.

\subsubsection{Near-horizon region}

In the near-horizon region $R \sim r_H$, the 2-point function has the form \eqref{eq:2pf-near}.
Substituting this into \eqref{eq:wigner-distribution}, we obtain
\beqa\label{eq:wd-near}
\mathcal{W}_1(x,p) &\approx& -\frac{3^3n_{2\beta}(p_0)}{2^5\pi^3p_0}
\left[1 + p_0P_1\Big(\frac{\mathbf{m}\mathbf{p}}{p}\Big)\partial_p
+ \frac{p_0^2}{3p}P_2\Big(\frac{\mathbf{m}\mathbf{p}}{p}\Big)
\big(p \partial_p^2 - \partial_p\big)\right]{\delta(p_0 - p)}\,,
\eeqa
where we have taken into account the condition $p_0 > 0$ and
\beqa
\mathbf{m} \equiv (1,0,0)
\eeqa
in the Fermi coordinate frame. Note that the structure of $\mathcal{W}_1(x,p)$ is the same as that of 
$\mathcal{W}_0(x,p)$ up to the total prefactor and the sign of the dipole-like term.

The distribution $\mathcal{W}_1(x,p)$ does not contain a contribution from
$W_M(x_F,x_F') \approx W_\text{HH}(x,x')$ which is an important part of the total 2-point function $W_1(x,x')$
as it provides the proper singularity structure for the Feynman propagator. The reason is that it has
no modes with the positive frequency.\footnote{Thus, this will contribute if we extend the allowed values
of $p_0$ from $-\infty$ to $+\infty$ in the formula \eqref{eq:wigner-distribution}, but the physics in terms
of particles then becomes obscure.} This is reasonable as the Wigner distribution of the vacuum
2-point function in Minkowski space is trivial as well.

However, there is a \emph{non}-vanishing contribution of $W_{M}(x,x')$ to the renormalised stress tensor if we approximate it by $W_\text{HH}(x,x')$.\footnote{Strictly speaking, $W_M(x,x')$ and
$W_\text{HH}(x,x')$ are not equal to each other near horizon in the Schwarzschild frame, unless one
sets $f'(r)/2 = \kappa$ in $W_M(x,x')$ for $r \sim r_H$.
This subtlety originates from the coordinate transformation to the local inertial frame near horizon and is 
inessential for the derivation of $\mathcal{W}_1(x,p)$.}
The renormalisation is performed by employing the point-splitting technique and making
the substitution 
\beqa
W_\text{HH}(x,x') &\rightarrow& W_\text{HH}(x,x') - H(x,x')
\eeqa
where $H(x,x')$ is the Hadamard parametrix (e.g., see~\cite{Moretti,Decanini&Folacci}). The parametrix
serves to cancel the singular part of $W_\text{HH}(x,x')$ in the coincidence limit $x' \rightarrow x$. It
also provides extra non-vanishing (geometrical) terms in $\langle T_\nu^\mu \rangle$ including the trace
aka conformal anomaly. Specifically, it was found in~\cite{Page} that
\beqa\label{eq:hh-nbhh}
\langle T_b^a \rangle_\text{HH} &\approx& - \frac{\kappa^4}{120\pi^2}\,\text{diag}(3,3,1,1)
\quad \text{for} \quad R \;\sim\; r_H 
\eeqa
This result for $\langle T_\nu^\mu\rangle_\text{HH}$ is close to that obtained by the numerical
calculations \cite{Candelas,Howard&Candelas}. This implies that the prefactor in front of the parameter $a_\omega$
appearing in the bi-scalar $\chi_\omega(\mathbf{x},\mathbf{x}')$ has actually to depend on the
spatial coordinates and asymptotically approach $1$ as faster as $1/R^4$ for $R \rightarrow \infty$.

To sum it up, the distribution $\mathcal{W}_1(x,p)$ is expected to give the stress tensor renormalised
as in~\cite{Candelas}, i.e. the \emph{relative} part of the total energy-momentum tensor
$\langle T_\nu^\mu \rangle$ with respect to $\langle T_\nu^\mu \rangle_M$. This seems to be
reasonable as only this part provides the crucial term resulting in the black-hole evaporation and, hence,
related to the Hawking particles.

\subsection{Energy-momentum tensor $\langle \hat{T}_{\mu\nu}(x) \rangle$}

Having the distribution function $\mathcal{W}(x,p)$, we are able to compute the energy-momentum
tensor $\langle \hat{T}_{\mu\nu} \rangle$ as its second moment with respect to the momentum $p_\mu$,
namely
\beqa\label{eq:emt}
\langle \hat{T}_{\mu\nu}(x) \rangle &=& {\int}d^4p \; p_\mu\,p_\nu \,\langle\hat{\mathcal{W}}(x,p)\rangle\,,
\eeqa
where the integration over $p_0$ is in the interval $(0,+\infty)$.

\subsubsection{Far-horizon region}

Employing our result for the Wigner distribution far away from the black hole, we obtain
\beqa
\langle \hat{T}_\nu^\mu \rangle &\approx&
\frac{1}{4\pi R^2}{\int}\frac{dp_0}{2\pi}\,\frac{p_0 \Gamma_{p_0}}{e^{\beta p_0} - 1}
\left[
\begin{array}{ccc}
+1 & +1\\
-1 & -1
\end{array}
\right] \quad \text{for} \quad R \;\gg\; r_H\,,
\eeqa
where the indices $\mu,\nu$ run over $\{t,r\}$ and the rest elements of $\langle \hat{T}_\nu^\mu \rangle$ vanish.
We have introduced $\Gamma_{p_0} = 27(p_0 M)^2$ which corresponds to the DeWitt approximation that we
have been employing throughout this paper. It is worth pointing out
that the $tt$-component of $\langle \hat{T}_\nu^\mu \rangle$ is due to the monopole-like term in~\eqref{eq:wd-far},
whereas its non-diagonal elements come from the dipole-like term in $\mathcal{W}_0(x,p)$ and the
$rr$-component of the stress tensor originates from the monopole- and quadrupole-like term of the Wigner
distribution. This result for the stress tensor $\langle \hat{T}_\nu^\mu \rangle$ given above is consistent
with~\cite{Christensen&Fulling,Candelas} and can also be directly obtained by using, e.g., the point-splitting
technique (see Appendix~\ref{app:emt} for some details).

It is a well-known fact that the energy density far away from the black hole is positive, i.e.
$\langle \hat{T}_t^t \rangle >0$, as well as its flux in the radial direction is also positive, i.e.
$\langle \hat{T}_t^r \rangle >0$. It implies that there is a positive energy flux from the black
hole. Thus, we re-derive this Hawking's discovery by use of the quantum kinetic approach.

It is tempting to define an \emph{effective} Wigner distribution as follows
\beqa\nonumber
\mathcal{W}_{\text{eff},0}(x,p) &=& \frac{\Gamma_{p_0}}{32\pi^3p_0^{3}R^2}\,n_\beta(p_0)
\sum_{l = 0}^{2}(2l + 1)P_l\Big(\frac{\mathbf{n}\mathbf{p}}{p}\Big)\delta(p_0 - p)
\\[1mm]
&\approx& \frac{1}{8\pi^2}\frac{\Gamma_{p_0}}{p_0^{3}R^2}\,n_\beta(p_0)\,\delta(p_\theta)\,
\delta(p_\phi)\,\delta(p_0 - p)\,,
\eeqa
where we have extended the finite summation over $l$ to the infinity and used a sum representation of
the delta function in terms of the spherical harmonics. Furthermore, we may define an effective 
one-particle distribution
\beqa
f_{\text{eff},0}(x,p) &=& \frac{1}{8\pi^2}\,\frac{\Gamma_{p_0}}{p_0^2R^2}\frac{1}{e^{\beta p_0} - 1}\,
\delta(p^\theta)\delta(p^\phi)\,,
\eeqa
which has already been introduced in~\cite{Emelyanov-16c} with slightly different notations. We shall
demonstrate below its usefulness.

\subsubsection{Near-horizon region}

Substituting $\mathcal{W}_1(x,p)$ into Eq.~\eqref{eq:emt}, we obtain
\beqa
\langle \hat{T}_b^a \rangle &\approx&
\frac{1}{\pi r_H^2}{\int}\frac{dp_0}{2\pi}\,\frac{p_0 \Gamma_{p_0}}{e^{2\beta p_0} - 1}
\left[
\begin{array}{cc}
-1 & +1 \\
-1 & +1
\end{array}
\right] \quad \text{for} \quad R \;\sim\; r_H\,,
\eeqa
where $a,b$ run over $\{t_F,x_F\}$ and the rest elements of $\langle \hat{T}_b^a \rangle$ are
suppressed in the Schwarzschild frame (see Appendix~\ref{app:emt} for further details).

The energy density $\langle \hat{T}_{t_F}^{t_F} \rangle$ is negative near the horizon, whereas
$\langle \hat{T}_{t_F}^{x_F} \rangle$ is positive. This physically implies that there is a flux of negative energy
towards the black hole. The change of the energy flux direction well away from the black-hole horizon
was first found in~\cite{Unruh} with the physical insight that the vacuum spacetime itself is unstable at the
quantum level. The same observation has been recently made in~\cite{Giddings}.

Analogous to the far-horizon region, one can introduce an effective Wigner distribution and associated with
it an effective one-particle distribution function, namely
\beqa\label{eq:eopdf-nhr}
f_{\text{eff},1}(x,p) &=& -\frac{1}{2\pi^2r_H^2}\,\frac{\Gamma_{p_0}}{e^{2\beta p_0} - 1}\,
\theta(-p_x)\delta(p_y)\delta(p_z)\,,
\eeqa
This correctly reproduces $\langle \hat{T}_b^a \rangle$ as well as $\langle N^a \rangle$ which we shall
compute below.

\subsection{Particle four-current $\langle \hat{N}_{\mu}(x) \rangle$}

In the kinetic theory, the first moment with respect to the four-momentum $p_\mu$ of the distribution
function $\mathcal{W}(x,p)$ gives the particle four-current. Specifically, we have
\beqa\label{eq:particle-current}
\langle \hat{N}_{\mu}(x) \rangle &=& {\int}d^4p \; p_\mu \;\langle\hat{\mathcal{W}}(x,p)\rangle
\quad \text{with} \quad p_0 \;\in\; (0,+\infty)\,.
\eeqa
Accordingly, the particle number density and its current are
\beqa
n(x) &=& \langle \hat{N}^{0}(x) \rangle \quad \text{and} \quad
N^i(x) \;=\; \langle \hat{N}^{i}(x)\rangle\,,
\eeqa
where the hydrodynamical velocity $U^\mu$ has been chosen of the form $(1,0,0,0)$. We merely
note that $U^\mu$ corresponds neither Eckart nor Landau-Lifshitz velocity as these have to be
light-like for the scalar model we have been considering.

We now go over to the study of this local macroscopic observable far away and close to the
black-hole horizon.

\subsubsection{Far-horizon region}

Substituting $\mathcal{W}_0(x,p)$ given in Eq.~\eqref{eq:wd-far} in the formula \eqref{eq:particle-current},
we obtain
\beqa
n_0(x) &=& N_0^r(x) \;=\;
\frac{1}{4\pi R^2}{\int}\frac{dp_0}{2\pi}\,\frac{\Gamma_{p_0}}{e^{\beta p_0} - 1}
+ \text{O}\Big(\frac{r_H^2T_H^4}{R^3}\Big)\,,
\eeqa
whereas $N_0^\theta(x) = N_0^\phi(x) = 0$ identically. Note that this result
can also be obtained with the aid of the effective one-particle distribution $f_{\text{eff},0}(x,p)$. It should also
be emphasised that $n_0(x)$ originates from the monopole-like term of $\mathcal{W}_0(x,p)$,
whereas $N_0^r$ comes from the dipole-like term in the Wigner distribution.

The number density as well as its current are positive, i.e. there is a positive radial particle flux
from the black hole. To better understand what this means we go over to the region near the horizon.

\subsubsection{Near-horizon region}

Substituting $\mathcal{W}_1(x,p)$ in the definition of the particle four-current, we obtain
\beqa
n_1(x) &=& -N_1^x(x) \;\approx\; -
\frac{1}{\pi r_H^2}{\int}\frac{dp_0}{2\pi}\,\frac{\Gamma_{p_0}}{e^{2\beta p_0} - 1}\,,
\eeqa
while $N_1^{y}(x)$ and $N_1^{z}(x)$ are zero.\footnote{We have suppressed the index
``$F$" in the Fermi coordinates for the sake of transparency of the formulas. This should
not cause any confusions as we employ all the time local inertial coordinates in both regions.}
This result implies that the density number of particles is \emph{negative} at $R \sim r_H$,
whereas its current is positive. The physical interpretation of $n_1(x) < 0$ in terms of particles
is here problematic as the number of particles per cubic centimetre cannot make any physical
sense whenever negative.

One of the possible explanation of this result might be that quantum kinetic theory cannot
adequately describe local physics near the horizon. Although the Wigner distribution
$\mathcal{W}_1(x,p)$ properly reproduces the evaporation effect of black holes, the particle
density may not have physical sense as the notion of particle may not be well-defined at
$R \sim r_H$. On the other hand, it seems that the Wigner's concept of particle should be
applicable in any local Minkowski frame, otherwise it would be unnatural to assume that
this concept holds in local (approximately) Minkowski frame on earth only. If the Wigner's
particle turns out to be physically realised at $R \sim r_H$, then $n_1(x) < 0$ has to be physically
understood.

Taking into account that there is no necessarily probabilistic interpretation of the Wigner
distribution for quantum systems~\cite{deGroot&vanLeeuwen&vanWeert},
it seems that there is still a physically non-excludable way of understanding $n_1(x) < 0$.
Specifically, one might think about $n_1(x)$ as a number of the field \emph{modes} per cubic
centimetre \emph{relative} to its number density in local Minkowski frame. If so, then $n_1(x) < 0$
would mean that the number of the field modes is smaller with respect to the flat case near the
horizon.\footnote{Note that if we consider a one-cubic-meter-size box with the gas of scalar particles
of temperature $T > T_H$, then $n_1(x)$ will be positive within the volume of this box. We explain 
below why $T$ must actually be much bigger than $T_H$, i.e. $T \gg T_H$, in order for this
set-up to make physical sense.} As a
consequence, $n_1(x) = -N_1^x(x) < 0$ should then imply the mode number decreases when
one approaches the horizon. If one also associates a positive energy $p_0$ with each
mode, one can then understand $\langle \hat{T}_{t_F}^{t_F} \rangle < 0$ as the total mode
energy density relative to their total energy density in the absence of the black hole. In other
words, this picture seems to fit well the near-horizon behaviour of the stress tensor
$\langle \hat{T}_b^a \rangle$ following from $\mathcal{W}_1(x,p)$.

This manner of interpreting $n_1(x) < 0$ as well as $\langle \hat{T}_{t_F}^{t_F} \rangle < 0$
is mostly motivated by the physical understanding of the Casimir effect. This viewpoint is also
consistent with our previous insights~\cite{Emelyanov-15b}. We come back to this issue below.

Comparing the behaviour of the particle four-current at $R \sim r_H$ and $R \gg r_H$, we find
that $n(x)$ changes its sign at a certain distance $R_c$ away from the event horizon. We
expect that it is of the order of $3M$, i.e. at the distance where the energy flux changes its sign (e.g.,
see~\cite{Giddings,Giddings-2,Hod,Dey&Liberati&Pranzetti,Giddings-3}). In one of our forthcoming papers, we
shall try to carefully study this region with the help of the particle four-current.

\subsection{Entropy four-current $\langle \hat{S}^\mu(x) \rangle$}

The macroscopic variables $\langle \hat{T}_\nu^\mu \rangle$ and
$\langle \hat{N}^\mu \rangle$ at $R \gg r_H$ behave like those of a steady flux of the
stellar wind of distance-independent temperature. Therefore, $s_0(x)$ coincides
with the entropy density of that kind of the idealised stellar wind.

As shown above, this picture is inapplicable in the near-horizon region. Moreover, the
entropy density $s_1(x)$ turns out to be imaginary. Specifically, its imaginary part is
ambiguous and reads
\beqa
\text{Im}\, s_1(x) &=& (\pi + 2\pi k)\,n_1(x) \quad \text{with} \quad k \;\in\; \mathbb{Z}\,.
\eeqa
We do not understand how it can be interpreted in terms of statistical properties of
some normal many-particle system.

\section{Concluding remarks}
\label{sec:cr}

\subsection{Scalar field splitting}

If we consider the \emph{fundamental} field operator $\hat{\Phi}(x)$ in the local inertial frame
in the far-horizon and near-horizon region, then we find that it possesses the following structure:
\beqa\label{eq:operator-sum}
\hat{\Phi}(x) &=& \hat{\Phi}_M(x) + \delta\hat{\Phi}(x) \quad \text{with} \quad
[\hat{\Phi}_M(x),\delta\hat{\Phi}(x')] \;=\; 0\,,
\eeqa
where $\hat{\Phi}_M(x)$ is the field operator as if there is no black hole, whereas $\delta\hat{\Phi}(x)$
vanishes as $r_H/R$ in the asymptotically flat region ($R \gg r_H$) and is of $\text{O}(1)$ near the
black-hole horizon ($R \sim r_H$). 
The field operator $\hat{\Phi}(x)$ before the collapse can be split in
a sum of two \emph{non}-fundamental operators with non-intersecting supports, namely $\hat{\Phi}_<(x)$
and $\hat{\Phi}_>(x)$, such that $\hat{\Phi}_<(x)$ vanishes for the Finkelstein-Eddington time $v > v_H$,
where $v_H$ corresponds to the moment when the event horizon forms, whereas $\hat{\Phi}_>(x)$
vanishes for $v < v_H$. One can further split $\hat{\Phi}_<(x)$ into $\hat{\Phi}_c(x)$ and $\hat{\Phi}_b(x)$~\cite{Hawking,DeWitt},
such that $\hat{\Phi}_c(x)$ has a vanishing support outside of the black-hole horizon, whereas $\hat{\Phi}_b(x)$
vanishes inside the horizon.\footnote{It should be noted that the modes $u_{in}(x|l,m,\omega,1)$ defined in~\cite{DeWitt}
are associated with the operator $\hat{\Phi}_<(x)$, while $u_{in}(x|l,m,\omega,2)$ with $\hat{\Phi}_>(x)$. The modes
$u_{in}(x|l,m,\omega,1)$ can be further split into $u_{out}(x|l,m,\omega,0)$ and $u_{out}(x|l,m,\omega,1)$.
These are related to $\hat{\Phi}_c(x)$ and $\hat{\Phi}_b(x)$, respectively.}

In terms of the non-fundamental operators $\hat{\Phi}_>(x)$ and $\hat{\Phi}_b(x)$ for $R > r_H$,
we have
\bsubeqs
\beqa
\hat\Phi_M(x) &=& \hat{\Phi}_>(x)
\quad\text{and}\quad 
\delta\hat\Phi(x) \;=\; \hat{\Phi}_b(x) \hspace{5.1mm} \text{for} \quad R \;\gg\; r_H\,,
\\[1mm]
\hat\Phi_M(x) &=& \hat{\Phi}_b(x)
\hspace{5.1mm}\text{and}\quad 
\delta\hat\Phi(x) \;=\; \hat{\Phi}_>(x) \quad \text{for} \quad R \;\sim\; r_H\,.
\eeqa
\esubeqs
Therefore, the operator $\hat{\Phi}_b(x)$ is as physically relevant as $\hat{\Phi}_>(x)$ and
\emph{vice verse} for having a proper singularity structure in the field propagator far away from as well as near
to the event horizon. It implies, for instance, that it is not legitimate to omit $\hat{\Phi}_>(x)$ in the
asymptotically flat region, contrary to the common practice. Precisely this part of $\hat{\Phi}(x)$ has been
successfully exploiting in particle physics, but do \emph{not} contribute to the covariant Wigner
distribution $\mathcal{W}_0(x,p)$.

The crucial role is, however, played by $\hat{\Phi}_b(x)$ near the event horizon as this part of
$\hat{\Phi}(x)$ provides the proper singularity in the 2-point function $W_1(x,x')$ and, hence, allows to
have the Feynman propagator with its ordinary interpretation in particle physics. The Wigner distribution
$\mathcal{W}_1(x,p)$ we have derived above is completely independent on $\hat{\Phi}_b(x)$.

The splitting~\eqref{eq:operator-sum} is of no physical sense in a local inertial frame falling in the black-hole
geometry. Still, the field operator $\hat{\Phi}(x)$ as being fundamental and its Hilbert space representation
makes physical sense all the way down to the black hole. This is contrary to the tacitly proposed idea to
define a separate Hilbert space for each of the non-fundamental operators on the right-hand side of
Eq.~\eqref{eq:operator-sum}. This idea eventually leads to the conclusion that the far-horizon region has to
be described by a thermal density matrix. We do not share this point of view as it is beyond of our current
understanding of local quantum field theory and, actually, inconsistent with that by
construction~\cite{Emelyanov-15b}.

\subsection{Scalar field particles and Wigner distribution}

The scalar operator $\hat{\Phi}(x)$ acquires the rich physical meaning in QFT when one represents it as
the sum of two non-Hermitian field operators, namely $\hat{\Phi}(x) = \hat{a}(x) + \hat{a}^\dagger(x)$. The 
operator $\hat{a}(x)$ is in turn defined through the equation
\beqa
\hat{a}(x) &=& {\int}\frac{d^4k}{(2\pi)^3}\,\theta(k_0)\delta(k^2)\,\Phi_\mathbf{k}(x)\,\hat{a}_\mathbf{k}\,,
\eeqa
where $\Phi_\mathbf{k}(x)$ are the mode functions being positive-frequency solutions (with respect to
$P_0$ of the \emph{local}\footnote{We find ourselves in a local (approximately)
inertial frame on earth.
Therefore, the Poincar\'{e} group in particle physics is local as well. Although the universe is not globally flat at
macroscopic scales, the Minkowski-space approximation is fully enough to successfully describe
scattering processes in the particle colliders.} Poincar\'{e} group) of the scalar field equation and satisfy
the normalisation condition $(\Phi_\mathbf{p},\Phi_\mathbf{k})_\text{KG} = \delta(\mathbf{p}-\mathbf{k})$.
The vacuum $|\Omega\rangle$ is defined through the equation $\hat{a}_\mathbf{k}|\Omega\rangle = 0$.

The one-particle state $|\mathbf{k}\rangle = \hat{a}_\mathbf{k}^\dagger|\Omega\rangle$ is not normalisable. 
The \emph{physical} 1-particle state is defined through the covariant wave packet $h(x)$:
\beqa
h(x) &=& {\int}\frac{d^4k}{(2\pi)^3}\,\theta(k_0)\delta(k^2)\,h(k)\Phi_\mathbf{k}(x)\,,
\eeqa
where $h(k)$ is a square-integrable function. A \emph{localised} particle state described by $h(k)$ is
\beqa
|h\rangle &\equiv& \hat{a}^\dagger(h)|\Omega\rangle
\;\equiv\; (h^*,\hat{\Phi})_\text{KG}|\Omega\rangle
\;=\; {\int}\frac{d^4k}{(2\pi)^3}\,\theta(k_0)\delta(k^2)\,h(k)|\mathbf{k}\rangle\,,
\eeqa
which is normalisable, i.e. $\langle h|h\rangle = 1$, as having a finite support.

There are infinitely many ways of splitting the field operator $\hat{\Phi}(x)$ in the sum of non-Hermitian
operators. This is a direct consequence of the linearity of the field equation. The proposal was to choose
different mode functions for different coordinate frames. This usually implies that it is meaningful to have
different notions of particles in different frames. This resulted eventually in a belief that ``quantum
mechanics is observer-dependent". We do not share this point of view as it leads to the various
paradoxical/unphysical conclusions. Recently, we have
proposed another principle which is conservative in its spirit and based on the idea of the equivalence
principle~\cite{Emelyanov-16c}. To make it short, the mode functions $\Phi_\textbf{p}(x)$ defining a
\emph{physical}, observer-\emph{in}dependent notion of particles are those which acquire the Minkowski
structure, namely
\beqa\label{eq:pm}
\Phi_\mathbf{k}(x) &\sim& \exp(-ik_\mu x^\mu)
\eeqa
in a local inertial frame defined at each point of spacetime. This makes sense only in space-time regions with
not too strong gravity. The main argument in favour of this definition is that we have been doing this all the
time on earth to predict and describe various scattering processes in the particle colliders.

Indeed, a well-tested notion of the particle is associated with the unitary, irreducible representations of the
Poincar\'{e} group $\mathcal{P}_+^\uparrow$. This idea was proposed long ago by Wigner (e.g.,
see~\cite{Haag}). The Poincar\'{e} group forms here the isometry of \emph{local} Minkowski frame only, as
the universe is globally non-flat. This is a basic idea behind of our proposal of relating the well-tested notion
of the particle in Minkowski space with its definition in curved spacetime. Note that the particle in a
non-inertial frame is described by an appropriate covariant wave packet of non-vanishing acceleration.

Once we have defined a wave packet, we have the 1-particle state carrying information about the
particle. The wave packet is characterised by its non-vanishing support. Normally, it should correspond 
to the size of the particle. In our case, this is given by the de Broglie wavelength $\lambda_\mathbf{k}$
of the scalar particle. Therefore, the correction to the right-hand side of \eqref{eq:pm} near
horizon must be suppressed by a factor of $(\lambda_\mathbf{k}/r_H)^2 \ll 1$, otherwise there is no
well-defined notion of the particle in the Wigner sense.\footnote{It seems that we are in agreement at this
point with~\cite{Bardeen} (see paragraph 3 on p. 2).} This is indeed the case.

Thus, we cannot relate the Wigner distribution $\mathcal{W}(x,p)$ we have found above to the \emph{real}
particles as this originates from the suppressed correction to the right-hand side of~\eqref{eq:pm}.

\subsection{Negative particle density and quantum noise}
 
The main idea of defining $\mathcal{W}(x,p)$ in QFT is to have a distribution function derived from
the first principles with the aid of which one can determine the local macroscopic state variables characterising
many-particle systems~\cite{deGroot&vanLeeuwen&vanWeert}. Indeed, we have seen that the Wigner
distribution $\mathcal{W}(x,p)$ can be used to compute the stress tensor $\langle \hat{T}_\nu^\mu\rangle$
and the particle four-flow $\langle \hat{N}^\mu\rangle$ as its second and first moment with respect to the
four-momentum $p_\mu$, respectively.

We have shown that the particle four-current
$N^\mu = (n_0,N_0^r,0,0)$ can make physical sense as a steady outward particle flow in the
asymptotically flat region. This is in full agreement with~\cite{Hawking}. However, this interpretation of
$N^\mu = (n_1,N_1^x,0,0)$ is inapplicable in the near-horizon region, because $n_1 < 0$ cannot be
possible for the real particles and qualitatively differs from a behaviour of a normal relativistic gas~\cite{Kremer}.

As pointed out above, $\mathcal{W}(x,p)$ comes in the present set-up from the correction to the leading
term of the mode functions (see Eq.~\eqref{eq:pm}). This correction plays a sub-leading role in the definition
of the particle creation operator $\hat{a}^\dagger(h)$ of the wave function $h(x)$, but the leading role
for $\mathcal{W}(x,p)$ to be non-trivial. Therefore, we think that $\mathcal{W}(x,p)$ with its moments 
are entirely due to the quantum fluctuations described by that correction which is in turn induced by the
presence of the black hole. The number of the modes characterising these fluctuations turns out to be
smaller at $R < R_c$ than that in the absence of the black hole. As a consequence, its relative number
density and energy density are negative.

If so, a noval property of the quantum fluctuations would be their ``ability" to transfer energy (through gravity
playing a role of the ``working body"). This does not seem to be a completely speculative idea bearing in mind
a lab set-up we described in~\cite{Emelyanov-16d}. Specifically, one can compute the vacuum energy density
in two cavities separated by an extra metallic plate in the Casimir set-up when this plate is in the middle and
when it is shifted in a way the dynamical Casimir effect is negligible.
Comparing the total vacuum energy density after and before the shift, one finds that its absolute value
has increased. Thus, the negative vacuum energy has been partially redistributed between the cavities and
partially dissipated in the middle plate by heating it up. The middle plate in this process plays a role of the
working body.

\section*{
ACKNOWLEDGMENTS}
It is a pleasure to thank Frans Klinkhamer and Jos\'{e} Queiruga for discussions.

\begin{appendix}
\section{Vacuum expectation value of stress tensor $\hat{T}_{\mu\nu}(x)$}
\label{app:emt}

The stress tensor $T_{\mu\nu}(x)$ of the (classical) massless scalar field $\Phi(x)$ conformally coupled to
gravity is given by
\beqa
T_{\mu\nu} &=& \frac{2}{3}\nabla_\mu\Phi\nabla_{\nu}\Phi - \frac{1}{6}\,g_{\mu\nu}
\nabla_\lambda\Phi \nabla^\lambda\Phi
-\frac{1}{3}\Phi\nabla_\mu\nabla_\nu\Phi\,.
\eeqa

Employing the point-splitting technique to get the renormalised value of the radial energy flux, we obtain
\beqa
\langle \hat{T}_{tr} \rangle &=& \frac{1}{3}{\lim_{x' \rightarrow x}}\Big[
(\partial_t\partial_{r'} + \partial_{t'}\partial_r)
- \frac{1}{2}(\nabla_t\nabla_r + \nabla_{t'}\nabla_{r'})\Big]W(x,x') \;=\; \pm
\frac{r_H^2}{r^2f}{\int}\frac{d\omega\,\omega}{(4\pi)^2}\,b_\omega\,
\eeqa
for both the far-horizon and near-horizon region.
The $\Delta{r}_\star$-term in the bi-scalar $\chi_\omega(\mathbf{x},\mathbf{x}')$ is crucial for having
non-vanishing radial energy flux.

It is straightforward to further show that
\bsubeqs
\beqa
\langle (\partial_t\hat{\Phi})^2\rangle &=&
+\frac{g_{tt}}{f}{\int}\frac{\omega d\omega}{(4\pi)^2}\,\bar{\chi}_\omega(\mathbf{x},\mathbf{x}),
\\[2mm]
\langle (\partial_i\hat{\Phi})^2\rangle &=&
-\frac{g_{ii}}{3f}{\int}\frac{\omega d\omega}{(4\pi)^2}
\left[1+\frac{\bar{R}_i^i}{2\omega^2f}\right]\bar{\chi}_\omega(\mathbf{x},\mathbf{x})
+{\int}\frac{d\omega}{(4\pi)^2\omega}
\lim_{\mathbf{x}' \rightarrow \mathbf{x}}\partial_{(i}\partial_{i')}\bar{\chi}_\omega(\mathbf{x},\mathbf{x}'),
\eeqa
\esubeqs
where there is no summation over $i = \{r,\theta,\phi\}$ in the second line, and we have introduced
a new bi-scalar as follows
\beqa
\bar{\chi}_\omega(\mathbf{x},\mathbf{x}') &\equiv&
\frac{\chi_\omega(\mathbf{x},\mathbf{x}')}{(f(r)f(r'))^\frac{1}{2}}\,,
\eeqa
and
\bsubeqs
\beqa
\langle \hat{\Phi}\hat{\Phi}_{;tt} \rangle &=& 
-\frac{g_{tt}}{f}{\int}\frac{\omega d\omega}{(4\pi)^2}\,\bar{\chi}_\omega(\mathbf{x},\mathbf{x})
+ \frac{1}{2}{\int}\frac{d\omega}{(4\pi)^2\omega}\lim_{r' \rightarrow r}
\big[\nabla_t^2{+}\nabla_{t'}^2\big]\bar{\chi}_\omega(r,r'),
\\[2mm]
\langle \hat{\Phi}\hat{\Phi}_{;ii} \rangle &=&
\frac{g_{ii}}{3f}{\int}\frac{\omega d\omega}{(4\pi)^2}
\left[1{+}\frac{\bar{R}_i^i}{2\omega^2f}\right]\bar{\chi}_\omega(\mathbf{x},\mathbf{x})
{+}\frac{1}{2}{\int}\frac{d\omega}{(4\pi)^2\omega}
\lim_{\mathbf{x}' \rightarrow \mathbf{x}}\big[\nabla_i^2{+}\nabla_{i'}^2\big]
\bar{\chi}_\omega(\mathbf{x},\mathbf{x'}).
\eeqa
\esubeqs
The vacuum expectation value of the trace of the non-renormalised stress tensor must vanish:
\beqa
\langle \hat{T}_\mu^\mu \rangle &=& -\frac{1}{3}
\langle \hat{\Phi}\Box\hat{\Phi} \rangle \;=\; -\frac{1}{3f^2}\int_\mathbf{R}\frac{d\omega}{(4\pi)^2\omega}\,
\lim_{\mathbf{x}' \rightarrow \mathbf{x}}\bar{\Box}\chi_\omega(\mathbf{x},\mathbf{x}') \;=\; 0\,.
\eeqa

\subsubsection{Far-horizon region}

Substituting the bi-scalar $\chi_\omega(\mathbf{x},\mathbf{x}')$ given in Eq.~\eqref{eq:bi-scalar}
with Eq.~\eqref{eq:fun-coefficients}, we
obtain
\beqa
\langle \hat{T}_\nu^\mu\rangle
&=& \frac{1}{f^2}{\int}\frac{d\omega\,\omega a_\omega}{(4\pi)^2}
\left[
\begin{array}{cc}
1 & 0 \\
0 & -\frac{1}{3}{\cdot}1_{3{\times3}} \\
\end{array}
\right]
+ \frac{1}{fR^2}{\int}\frac{d\omega\,\omega b_\omega}{(4\pi)^2}
\left[
\begin{array}{ccc}
+1 & +1 & 0 \\
-1 & -1 & 0\\
0 & 0 & 0_{2{\times}2} \\
\end{array}
\right] +\text{O}\Big(\frac{1}{R^5}\Big)
\eeqa
in the far-horizon region, i.e. for $R \gg r_H$.

\subsubsection{Near-horizon region}

In the near-horizon region, i.e. $R \sim r_H$, the non-renormalised stress-energy tensor reads
\beqa\label{eq:emt-nhr}
\langle \hat{T}_\nu^\mu\rangle
&=& \frac{1}{f^2}{\int}\frac{d\omega\,\omega a_\omega}{(4\pi)^2}
\left[
\begin{array}{cc}
1 & 0 \\
0 & -\frac{1}{3}{\cdot}1_{3{\times3}} \\
\end{array}
\right]
+ {\int}\frac{d\omega\,\omega b_\omega}{(4\pi)^2r_H^2}\left[
\begin{array}{ccc}
+\frac{1}{f} & - 1 & 0 \\
+\frac{1}{f^2} & -\frac{1}{f} & 0\\
0 & 0 & 0_{2{\times}2} \\
\end{array}
\right]
\\[1mm]\nonumber
&&\hspace{20mm} -\frac{1}{3r_H^4}{\int}\frac{d\omega\, b_\omega}{(4\pi)^2\omega}
\left[
\begin{array}{cc}
+1_{2{\times}2} & 0 \\
0 & -1_{2{\times}2} \\
\end{array}
\right]{\times}\;\Big(1 -\frac{11}{2}f + \frac{35}{2}f^2+\text{O}\big(f^3\big)\Big)
\eeqa
for $\alpha_\omega(r,r')$ and $\beta_\omega(r,r')$ given in~\eqref{eq:fun-coefficients}.

The matrix structure of the third term in Eq.~\eqref{eq:emt-nhr} is of the crucial importance, because
it guarantees that this term is also finite on the horizon in the Fermi frame. It should
be noted that the extra corrections to $\alpha_\omega(r,r')$ and $\beta_\omega(r,r')$ also contribute 
to this term to the leading order changing its numerical value and the sign as follows from
the numerical results of~\cite{Elster}. This contribution to the stress tensor does
not change its value and structure when rewritten in the Fermi frame like the Hartle-Hawking part
(given in Eq.~\eqref{eq:hh-nbhh}).

The terms vanishing as $f(R)$ in the Schwarzschild frame also contribute in the Fermi frame.
It implies that the difference
$\langle \Delta\hat{T}_b^a\rangle \equiv \langle \hat{T}_b^a\rangle - \langle \hat{T}_b^a\rangle_\text{HH}$
in the Fermi frame is actually given by
\beqa\label{eq:delta-emt-nhr}
\langle \Delta\hat{T}_b^a\rangle &\approx& - \frac{L}{16\pi r_H^2}
\left[
\begin{array}{ccc}
+1 & - 1 & 0 \\
+1 & -1 & 0\\
0 & 0 & 0_{2{\times}2} \\
\end{array}
\right]
-{\gamma_1}\left[
\begin{array}{cc}
+1_{2{\times}2} & 0 \\
0 & -1_{2{\times}2} \\
\end{array}
\right]
-{\gamma_2}\left[
\begin{array}{ccc}
+1 & +1 & 0 \\
-1 & -1 & 0\\
0 & 0 & 0_{2{\times}2} \\
\end{array}
\right]
\eeqa
near the event horizon with
\bsubeqs
\beqa
\gamma_1 &=& \sum_{l = 0}^{+\infty}{\int}\frac{dx}{4\pi x}
\frac{(2l{+}1)|B_{\omega l}|^2}{e^{8\pi x}-1}\,\frac{l(l{+}1)(1 + 24x^2)+8x^2}
{6\pi r_H^4(1+16x^2)}\,,
\\[1mm]
\gamma_2 &=& \sum_{l = 0}^{+\infty}{\int}\frac{dx}{4\pi x}\frac{(2l{+}1)|B_{\omega l}|^2}{e^{8\pi x}-1}\,
\frac{2l(l{+}1)(1+40x^2)+3[l(l{+}1)]^2(1+8x^2) +72x^2}
{12\pi r_H^4(1+20x^2 +64x^4)}\,,
\eeqa
\esubeqs
where $x \equiv \omega M$, with the numerical values $\gamma_1 \approx 1.25{\times}10^{-6}/M^{4}$
(in agreement with~\cite{Elster}) and $\gamma_2 \approx 4.45{\times}10^{-6}/M^{4}$. It should be noted
that $L/16\pi r_H^2 \approx 9.25{\times}10^{-8}/M^{4}$ which is much smaller than $\gamma_2$. Still,
the decrease of the black-hole mass $M$ is entirely due to $L$, namely $\dot{M} = -L$, where dot stands for
the differentiation with respect to the Schwarzschild time coordinate. This means
that the last term in Eq.~\eqref{eq:delta-emt-nhr} is geometrical and it might be that one should throw
away this from the solution of the field equation. Note that this term
vanishes as $f(R)$ in the near-horizon region in the Schwarzschild frame.

The last two terms in~\eqref{eq:delta-emt-nhr} cannot be described by any one-particle distribution
$\tilde{f}_{\text{eff},1}(\mathbf{p})$, because it must satisfy the condition 
\beqa
{\int}\frac{d^3\mathbf{p}}{|\mathbf{p}|}\,\tilde{f}_{\text{eff},1}(\mathbf{p}) &=& 0\,.
\eeqa
This follows from 
\beqa
\Delta\tilde{W}_1(x,x') &=& \frac{1}{2}
\big((\gamma_1+\gamma_2)\Delta{t}^2 -2\gamma_2\Delta{t}\Delta{x}
- (\gamma_1-\gamma_2)\Delta{x}^2 + \gamma_1\Delta{y}^2 + \gamma_1\Delta{z}^2 \big)\,,
\eeqa
which vanishes in the coincidence limit, i.e. $x' \rightarrow x$. Note that 
$\Box \Delta\tilde{W}(x,x') = 0$ exactly holds and $\Delta\tilde{W}_1(x,x')$ is locally suppressed
as $(\Delta{x}/r_H)^2$ with respect to $\Delta{W}_1(x,x')$ and as $(\Delta{x}/r_H)^4$
with respect to $W_M(x,x') \approx W_\text{HH}(x,x')$.

\end{appendix}

\end{document}